\documentclass[aps,prl,twocolumn,superscriptaddress,showpacs,floatfix,preprintnumbers,amsmath,amssymb]{revtex4}

\usepackage{epsfig}

\RequirePackage{xspace}

\usepackage{relsize}

\def\babar{\mbox{\slshape B\kern-0.1em{\smaller A}\kern-0.1em
    B\kern-0.1em{\smaller A\kern-0.2em R}}}
\def\invfb   {\ensuremath{\mbox{\,fb}^{-1}}\xspace}
\def\ccbar {\ensuremath{c\overline c}\xspace}
\def\pep2{PEP-II}
\def\BB      {\ensuremath{B\Bbar}\xspace} 
\def\Bbar    {\kern 0.18em\overline{\kern -0.18em B}{}\xspace}
\def\iowaBranchingFractionProduct{\mathcal{B}(B \rightarrow \Xi_c^0 X) \times \mathcal{B}(\Xi_c^0 \rightarrow \Xi^- \pi^+)}
\def\iowaCrossSectionProduct{\sigma(e^+ e^- \rightarrow \Xi_c^0 X) \times \mathcal{B}(\Xi_c^0 \rightarrow \Xi^- \pi^+)}

\def\iowaBresult{2.11}
\def\iowaBstaterr{0.19}
\def\iowaBsyserr{0.25}
\def\iowaCresult{388}
\def\iowaCstaterr{39}
\def\iowaCsyserr{41}

\begin{document}

\preprint{\babar-PUB-05/008}
\preprint{SLAC-PUB-11100}

\title{Production and decay of $\Xi_c^0$ at \babar}


% \author{The \babar\ collaboration}
%\thanks{}
%\affiliation{}


%%%%%%%%%%%%%%%%%%%%%%% begin authors_pub05008_edited.tex %%%%%%%%%%%%%%%%%%%%%%%
%% author list as of 02-Feb-2005 (631 authors)
%
\author{B.~Aubert}
\author{R.~Barate}
\author{D.~Boutigny}
\author{F.~Couderc}
\author{Y.~Karyotakis}
\author{J.~P.~Lees}
\author{V.~Poireau}
\author{V.~Tisserand}
\author{A.~Zghiche}
\affiliation{Laboratoire de Physique des Particules, F-74941 Annecy-le-Vieux, France }
\author{E.~Grauges}
\affiliation{IFAE, Universitat Autonoma de Barcelona, E-08193 Bellaterra, Barcelona, Spain }
% ...
\collaboration{The \babar\ Collaboration}
\noaffiliation

%%%%%%%%%%%%%%%%%%%%%%%% end authors_pub05008_edited.tex %%%%%%%%%%%%%%%%%%%%%%%%

\date{\today}

\begin{abstract}
  Using 116.1~\invfb of data collected by the \babar\ detector,
  we present an analysis of $\Xi_c^0$ production in $B$
  decays and from the \ccbar continuum,
  with the $\Xi_c^0$ decaying into $\Omega^- K^+$ and $\Xi^- \pi^+$
  final states. We measure the ratio of branching fractions
  $\mathcal{B}(\Xi_c^0 \rightarrow \Omega^- K^+)/\mathcal{B}(\Xi_c^0 \rightarrow \Xi^- \pi^+)$
  to be $0.294 \pm 0.018 \pm 0.016$, where the first uncertainty
  is statistical and the second is systematic.
  The $\Xi_c^0$ momentum spectrum is measured on and 40~MeV below
  the $\Upsilon(4S)$ resonance. From these spectra the branching
  fraction product
  $\iowaBranchingFractionProduct$ is measured to be
  $(\iowaBresult \pm \iowaBstaterr \pm \iowaBsyserr) \times 10^{-4}$,
  and the cross-section product $\iowaCrossSectionProduct$
  from the continuum is measured to be
  $(\iowaCresult \pm \iowaCstaterr \pm \iowaCsyserr)$~fb
  at a center-of-mass energy of 10.58~GeV.
\end{abstract}

%% 14.20.Lq Charmed baryons
%% 13.30.Eg Hadronic decays of baryons
%% 13.25.Hw Hadronic decays of bottom mesons

%\keywords{}

\maketitle


In this Letter we present a study of the $\Xi_c^0 (csd)$~\cite{ref:PDGbook}
charmed baryon through two decay modes:
$\Xi_c^0 \rightarrow \Omega^-\:   K^+$ and
$\Xi_c^0 \rightarrow \Xi^-   \: \pi^+$~\cite{footnote:cc},
the former of which is expected to proceed almost entirely via internal W-exchange.
We determine the ratio of branching fractions of these decay modes,
which has been measured previously to be
$0.50 \pm 0.21 \pm 0.05$~\cite{Henderson:1992cx,footnote:errors}.
It was predicted to be 0.32 with a quark model calculation in which
no spin information is exchanged between quarks other than through
a single $W$ boson~\cite{Korner:1992wi}.

We also study $\Xi_c^0$ production by measuring the
spectrum of the $\Xi_c^0$ momentum in the 
$e^+ e^-$ center-of-mass frame ($p^*$).
A number of theoretical predictions for $\Xi_c$ production
in $B$ decays have been made~\cite{Ball:1990fw,Chernyak:1990ag,Sheikholeslami:1991fa,Dunietz:1994hj}.
There are several possible production mechanisms, principally
$b \rightarrow c \bar{c} s$ weak decays, $b \rightarrow c \bar{u} d$
weak decays in which an $s\bar{s}$ pair is produced during
fragmentation, and Cabibbo-suppressed $b \rightarrow c \bar{u} s$
weak decays. At this point there is insufficient experimental
evidence to determine which of these is the dominant mechanism,
and no clear theoretical consensus.
Insight into the contributing processes can be gained by studying
the shape of the $p^*$ spectrum. Evidence for $\Xi_c$ production
in $B$ decays was presented previously by the CLEO collaboration,
with a statistical significance of $\sim3\sigma$ in the
$\Xi_c^0 \rightarrow \Xi^- \pi^+$ decay mode and
$\sim4\sigma$ in the $\Xi_c^+ \rightarrow \Xi^- \pi^+ \pi^+$
decay mode~\cite{Barish:1997pq}.

The data for this analysis were collected with the \babar\ detector at
the SLAC \pep2\ asymmetric energy $e^+e^-$ collider; the detector is described
in detail elsewhere~\cite{ref:babar}.
A total integrated luminosity of
116.1~\invfb is used, of which 105.4~\invfb was collected at the
$\Upsilon(4S)$ resonance~\cite{ref:PDGbook} (corresponding to 116~million
\BB pairs) and 10.7~\invfb was collected at a
center-of-mass energy of 10.54~GeV, which is below the \BB 
production threshold. These are referred to as the on-resonance
and off-resonance data samples, respectively.

The reconstruction of $\Xi_c^0$ candidates takes place as follows.
A $\Lambda$ candidate is reconstructed by identifying a proton 
and combining it with an oppositely charged track interpreted
as a $\pi^-$, fitting the tracks to a common vertex.
The $\Lambda$ candidate is then combined with a negatively charged
track interpreted as a $\pi^-$ ($K^-$) to form a
$\Xi^-$ ($\Omega^-$) candidate. For each intermediate hyperon,
the invariant mass is required to be within 3$\sigma$ of
the central value, where $\sigma$ is the fitted mass
resolution. The invariant mass is then constrained to the
nominal value~\cite{ref:PDGbook}.
Each resulting $\Xi^-$ ($\Omega^-$) candidate passing the selection
criteria is then combined with a positively charged track
interpreted as a $\pi^+$ ($K^+$) to form a $\Xi_c^0$ candidate.
For the $\Omega^- K^+$ final state, the two $K^{\pm}$ tracks
must be identified as kaons. Particle identification is
performed with $dE/dx$ and Cherenkov angle measurements~\cite{ref:babar}.

\begin{figure}
  \begin{center}
    \epsfig{file=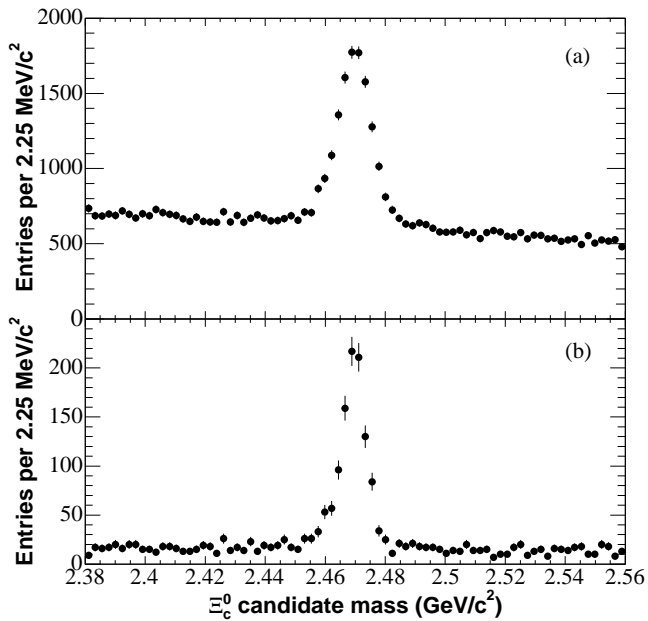, width=\columnwidth}
  \end{center}
  \caption[Reconstructed $\Xi_c^0$ spectra]
	  {Invariant mass distributions for $\Xi_c^0$
	    candidates in 116.1~\invfb of data, for
	    (a) $\Xi^- \pi^+$, and
	    (b) $\Omega^- K^+$.
	  }
  \label{fig:Xicdata}
\end{figure}

Additional selection criteria, described below, are used to
improve the signal-to-background ratio. As a precaution against
selection bias, these are optimized with
subsamples of the data: 20~\invfb and 40~\invfb
for the $\Xi^- \pi^+$ and $\Omega^- K^+$
final states, respectively. A minimum decay distance
of 2.5~mm (1.5~mm) between the event primary vertex and the
$\Xi^-$ ($\Omega^-$) decay vertex in the plane perpendicular to
the beam direction is required.
The distance between the $\Omega^-$ and $\Lambda$ decay
vertices is required to be at least 3~mm. In addition, the
relative positioning of vertices is required to be causally
connected: we reject candidates in which the $\Xi^-$ decays
further from the primary vertex than its daughter $\Lambda$ does,
or where the displacement vector from the $\Omega^-$ decay
point to the $\Lambda$ decay point is anti-parallel to the
$\Lambda$ momentum vector~\cite{footnote:flightcuts}.
The invariant mass distributions for the $\Xi_c^0$ candidates
in the full data set
satisfying these criteria are shown in
Fig.~\ref{fig:Xicdata}~(a) and~(b) for $\Xi^-\pi^+$ and
$\Omega^- K^+$ combinations, %respectively.
with signal yields of approximately
8100 and 1000 events, respectively.

Simulated events with the $\Xi_c^0$ decaying into the two
desired final states are generated for the processes
  $e^+e^- \rightarrow \ccbar \rightarrow \Xi_c^0 X$ and
  $e^+e^- \rightarrow \Upsilon(4S) \rightarrow \BB \rightarrow \Xi_c^0 X$,
where $X$ represents the rest of the event.
The \textsc{pythia} simulation package~\cite{ref:pythia}, tuned to the global \babar\ data,
is used for the $\ccbar$ fragmentation
and for $B$ decays to $\Xi_c^0$, and \textsc{geant4}~\cite{ref:geant4} is used
to simulate the detector response.
For \ccbar production,
samples of 90,000 events for the $\Xi^-\pi^+$ final state and 60,000
for the $\Omega^- K^+$ final state are generated. For \BB
production, samples of 255,000 and 120,000 events are used, respectively.

Additional generic Monte Carlo events are used
to investigate possible background contributions.
The sample sizes are
equivalent to 245~\invfb, 64~\invfb, and 33~\invfb for 
$e^+ e^- \rightarrow \BB$, $\ccbar$, and $q\bar{q}$,
respectively, where $q = {u,d,s}$.
Excluding signal contributions, the mass distribution
varies smoothly throughout the region near
the $\Xi_c^0$ mass.

To measure the ratio of branching fractions,
a further requirement that $p^* > 1.8$~GeV$/c$ is imposed
on the $\Xi_c^0$ candidates
in order to suppress combinatoric background and
improve the signal purity.
Additionally, the candidates are required to be within
the region of high $\Xi_c^0$ reconstruction efficiency
$-0.8 \leq \cos \theta^* \leq 0.6$, where
$\theta^*$ is the polar angle of the $\Xi_c^0$ candidate with
respect to the collision axis in the center-of-mass frame. 
After these criteria, the signal yields for the $\Xi^-\pi^+$
and $\Omega^- K^+$ modes are approximately 3650 and 650, respectively.
The efficiency is calculated from signal Monte Carlo events as a
function of  $p^*$ and $\cos\theta^*$ for each of the decay modes.
For each mode, a fifteen-parameter fit gives a smooth
parameterization of the efficiency with small statistical
uncertainty. The efficiency is then corrected by weighting
each candidate by the inverse of its efficiency,
and the efficiency-corrected mass
spectrum is fitted with a double Gaussian
with a common mean
for signal plus a linear background function. Including efficiency
loss due to the $\Omega^-$ and $\Lambda$ branching fractions,
we obtain 
$25889 \pm 516$ weighted events in the $\Xi^-\pi^+$ mode and
$7615 \pm 443$ weighted events in the $\Omega^- K^+$ mode.
The $\chi^2$ fit probabilities are 65\% and 5\%, respectively.
In each case, the wider Gaussian contributes approximately
one quarter of the yield.

We evaluate several sources of systematic uncertainty in the
ratio of branching fractions:
  the fits to the mass spectra (3.4\%),
  the efficiency parameterization (3.1\%),
  particle identification (2.0\%),
  finite Monte Carlo statistics (1.4\%),
  multiple candidates in the same event (1.0\%),
  charge asymmetries in detection efficiency (1.0\%),
  the $\cos\theta^*$ distribution (1.0\%), and
  the $\Omega^-$ branching fraction (1.0\%).
No baryon polarization is considered and any
systematic uncertainty due to this is neglected.
Adding all of the uncertainties in quadrature, we obtain:
\begin{displaymath}
  \frac{ \mathcal{B}(\Xi_c^0 \rightarrow \Omega^- K^+)}{\mathcal{B}(\Xi_c^0 \rightarrow \Xi^- \pi^+) } 
  =  0.294 \pm 0.018 \pm 0.016.
\end{displaymath}

After obtaining the ratio of branching fractions, we next
measure the $p^*$ spectrum of the $\Xi_c^0$ baryons
in order to study the production mechanisms in both \ccbar
and \BB events. The same selection criteria and data samples
described above are used, except that no requirement on
$p^*$ or $\cos\theta^*$ is made. Instead, the $\Xi_c^0$
candidates are divided into intervals of $p^*$. The yield is then
measured in each interval with two different methods:
first with a fitting method, where the mass spectrum
is fitted with a single Gaussian
for signal plus a linear background function and
the integral of the Gaussian is taken as the yield;
second with a counting method, where the background is estimated from mass sidebands
and the signal yield is then taken
as the statistical excess above this background
in a mass window around the peak.
The use of two different methods serves as a cross-check.

The efficiency in each $p^*$ interval is estimated with
signal Monte Carlo events from that $p^*$ range. For both methods,
the simulated events are reconstructed and the yield is measured,
then divided by the number of events generated to obtain the efficiency.
Due to the different angular distributions,
the efficiencies
for $\Xi_c^0$ produced from \ccbar $(\varepsilon_{\ccbar})$ 
and from \BB $(\varepsilon_{\BB})$ differ slightly.
In the region $1.2 < p^* < 2.0$~GeV$/c$ where both
production mechanisms are significant and the difference is
approximately 8\% (relative), the efficiency is
taken to be $(\varepsilon_{\ccbar} + \varepsilon_{\BB})/2$.
The systematic uncertainty on the efficiency is then
$|\varepsilon_{\ccbar} - \varepsilon_{\BB}|/\sqrt{12}$.
The angular distributions produced in \textsc{pythia} fragmentation are assumed
to be correct when calculating the efficiency; the data are fully
consistent with these distributions within available statistics.
The efficiency-corrected
yield in each $p^*$ interval is then calculated, including
loss of efficiency due to the $\Lambda$ and $\Omega^-$ branching
fractions.
The spectra obtained with the two methods are 
in good agreement; we use the counting method for the quoted
results since it is more stable for low statistics.

\begin{figure}
  \begin{center}
    \epsfig{file=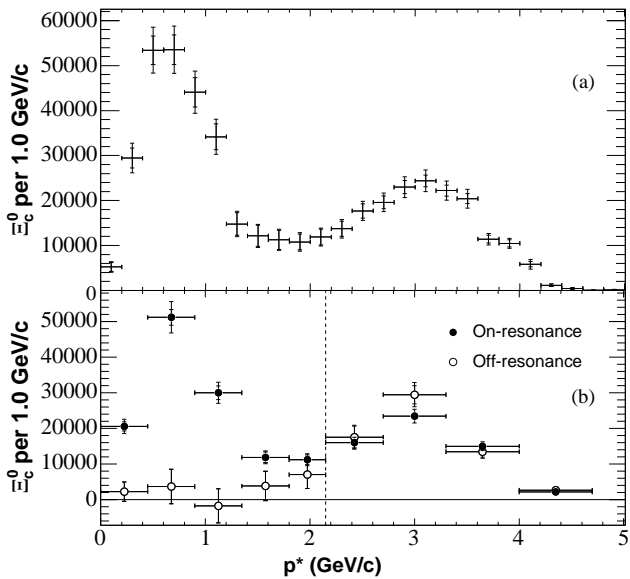, width=\columnwidth}
  \end{center}
  \caption[$p^*$ spectra]
	  {The $p^*$ spectrum measurements.
	    In (a), the $p^*$ spectrum of $\Xi_c^0$ decaying via $\Xi^- \pi^+$
	    is shown for the on-resonance data sample.
	    In (b), the on-resonance and off-resonance data samples are shown together,
	    with the off-resonance normalization scaled to account for the
	    difference in integrated luminosity and cross-section. 
	    In each plot, the inner error bars give the statistical
	    uncertainty and the outer error bars give the sum in quadrature
	    of the statistical and systematic uncertainties. The vertical
	    line at 2.15~GeV$/c$ in (b) shows the kinematic cutoff for $\Xi_c^0$ produced
	    in $B$ decays at \babar.
	    Note that the vertical axes show events per unit $p^*$, not events
	    in each $p^*$ bin as given in 
	    Table~\ref{tab:pstar:spectrum}.
	  }
  \label{fig:pstar}
\end{figure}

A number of systematic uncertainties are considered, the most
important of which are the uncertainties associated with the track-finding
and particle identification efficiencies (5.8\% and 3.5\%, respectively).
Uncertainties from 
  the simulated $\Xi_c^0$ mass resolution (1\%),
  the mass resolutions of the intermediate hyperon states (0.5\%), 
  the $p^*$ resolution [($\mathcal{O}(1\%)$)],
  the effect of finite interval width [($\mathcal{O}(2\%)$)], 
  multiple candidates (0\%),
  non-linearity of the background [($\mathcal{O}(1\%)$)],
  the signal measurement method used (2\%),
  the finite Monte Carlo statistics available [($\mathcal{O}(3\%)$)],
  and uncertainties in the $\Lambda$ and $\Omega^-$ branching fractions (0.8\%, 1.0\%)
are all considered individually; the notation $\mathcal{O}(x\%)$
indicates the typical value when the exact uncertainty varies
among $p^*$ intervals.
The total systematic uncertainty for each $p^*$ interval is
obtained by adding the individual contributions in quadrature.
In addition, a systematic
correction of 1.0\% is applied to account for a known
data-Monte Carlo discrepancy in the track-finding efficiency,
and small corrections are applied to each interval
to account for the broadening effect of the
$p^*$ experimental resolution on the spectrum.
The final $p^*$ spectrum for the on-resonance data set,
obtained with the counting method in the $\Xi^- \pi^+$ mode,
is shown in Fig.~\ref{fig:pstar}(a).
Table~\ref{tab:pstar:spectrum} shows the corresponding values.

\begin{table}
  \caption[Yield and cross-section by $p^*$ interval]
	  {
	    Efficiency-corrected yield
	    and cross-section product including $B$ production
	    $\iowaCrossSectionProduct$, for the on-resonance data.
	  }
  \begin{center}
    \begin{tabular}{cccccc}
      \hline \hline
      $p^*$ range && Corrected && Cross-section
   \\   (GeV$/c$) &&   yield   && product (fb)
      \\ \hline
      %% \input PRL_thintable-edited
%%%%%%%%%%%%%%%%%%%%%%% begin PRL_thintable-edited.tex %%%%%%%%%%%%%%%%%%%%%%%
       0.0--0.2 && $1046 \pm 201 \pm 128$  && $10 \pm 2 \pm 1$
\\     0.2--0.4 && $5889 \pm 446 \pm 483$  && $56 \pm 4 \pm 5$
\\     0.4--0.6 && $10681 \pm 631 \pm 801$ && $101 \pm 6 \pm 8$
\\     0.6--0.8 && $10709 \pm 660 \pm 817$ && $102 \pm 6 \pm 8$
\\     0.8--1.0 && $8811 \pm 647 \pm 679$  && $84 \pm 6 \pm 7$
\\     1.0--1.2 && $6834 \pm 573 \pm 530$  && $65 \pm 5 \pm 5$
\\     1.2--1.4 && $2954 \pm 501 \pm 252$  && $28 \pm 5 \pm 2$
\\     1.4--1.6 && $2429 \pm 470 \pm 212$  && $23 \pm 4 \pm 2$
\\     1.6--1.8 && $2252 \pm 424 \pm 202$  && $21 \pm 4 \pm 2$
\\     1.8--2.0 && $2159 \pm 350 \pm 217$  && $20 \pm 3 \pm 2$
\\     2.0--2.2 && $2375 \pm 347 \pm 205$  && $23 \pm 3 \pm 2$
\\     2.2--2.4 && $2743 \pm 340 \pm 227$  && $26 \pm 3 \pm 2$
\\     2.4--2.6 && $3537 \pm 315 \pm 285$  && $34 \pm 3 \pm 3$
\\     2.6--2.8 && $3920 \pm 282 \pm 306$  && $37 \pm 3 \pm 3$
\\     2.8--3.0 && $4595 \pm 294 \pm 359$  && $44 \pm 3 \pm 3$
\\     3.0--3.2 && $4873 \pm 263 \pm 401$  && $46 \pm 2 \pm 4$
\\     3.2--3.4 && $4442 \pm 244 \pm 348$  && $42 \pm 2 \pm 3$
\\     3.4--3.6 && $4084 \pm 223 \pm 355$  && $39 \pm 2 \pm 3$
\\     3.6--3.8 && $2282 \pm 171 \pm 189$  && $22 \pm 2 \pm 2$
\\     3.8--4.0 && $2095 \pm 155 \pm 166$  && $20 \pm 1 \pm 2$
\\     4.0--4.2 && $1168 \pm 123 \pm 177$  && $11 \pm 1 \pm 2$
\\     4.2--4.4 && $233 \pm 53 \pm 32$     && $2.2 \pm 0.5 \pm 0.3$
\\     4.4--4.6 && $88 \pm 37 \pm 21$      && $0.8 \pm 0.3 \pm 0.2$
\\     4.6--4.8 && $5 \pm 13 \pm 7$        && $0.0 \pm 0.1 \pm 0.1$
\\     4.8--5.0 && $24 \pm 17 \pm 16$      && $0.2 \pm 0.2 \pm 0.1$
%%%%%%%%%%%%%%%%%%%%%%%% end PRL_thintable-edited.tex %%%%%%%%%%%%%%%%%%%%%%%%
      \\ \hline \hline
    \end{tabular}
  \end{center}
  \label{tab:pstar:spectrum}
\end{table}

\begin{table}
  \caption[Cross-section by $p^*$ interval]
	  {
	    Cross-section product including
	      $B$ production $\iowaCrossSectionProduct$,
	      for the on- and off-resonance data. The off-resonance
	      cross-sections are scaled to a center-of-mass energy
	      of 10.58~GeV.
	  }
  \begin{center}
    \begin{tabular}{cccccc}
      \hline \hline
      $p^*$ range && \multicolumn{3}{c}{Cross-section product (fb)}
   \\   (GeV$/c$) && On-resonance && Off-resonance
      \\ \hline
      %%\input PRL_thintable_offpeak-edited
%%%%%%%%%%%%%%%%%%%%%%% begin PRL_thintable_offpeak-edited.tex %%%%%%%%%%%%%%%%%%%%%%%
       0.00--0.45 && $ 88 \pm 5 \pm  7$ && $ 10 \pm 12 \pm  1$
\\     0.45--0.90 && $218 \pm 9 \pm 17$ && $ 16 \pm 21 \pm  2$
\\     0.90--1.35 && $128 \pm 8 \pm 10$ && $ -7 \pm 20 \pm  2$
\\     1.35--1.80 && $ 51 \pm 6 \pm  4$ && $ 16 \pm 18 \pm  2$
\\     1.80--2.15 && $ 37 \pm 4 \pm  3$ && $ 23 \pm 13 \pm  2$
\\     2.15--2.70 && $ 83 \pm 5 \pm  6$ && $ 91 \pm 16 \pm  7$
\\     2.70--3.30 && $133 \pm 4 \pm 10$ && $168 \pm 15 \pm 13$
\\     3.30--4.00 && $ 99 \pm 3 \pm  8$ && $ 89 \pm 10 \pm  7$
\\     4.00--4.70 && $ 14 \pm 1 \pm  1$ && $ 17 \pm  4 \pm  2$
%%%%%%%%%%%%%%%%%%%%%%%% end PRL_thintable_offpeak-edited.tex %%%%%%%%%%%%%%%%%%%%%%%%
      \\ \hline \hline
    \end{tabular}
  \end{center}
  \label{tab:pstar:spectrum-offpeak}
\end{table}

A further check is performed by comparing
the two decay modes. The $\Omega^- K^+$ yields are scaled
by a factor of $(1/0.294)$, the ratio of branching
fractions previously presented in this Letter.
Because the $\Omega^- K^+$ signal has
fewer events, wider $p^*$ intervals are used. The spectra of the
two modes show good agreement in both shape and normalization
and have a $\chi^2$ probability of 80\% for consistency.
This serves as a cross-check both of the $p^*$ spectrum
measurement and of the ratio of branching fractions.

The double-peak structure seen in the $p^*$ spectrum is
due to two production mechanisms: the peak at lower
$p^*$ is due to $\Xi_c^0$ production in $B$ meson decays
and the peak at higher $p^*$ is due to $\Xi_c^0$ production
from the \ccbar continuum. This is evident from 
Fig.~\ref{fig:pstar}(b), where the $p^*$ spectra for 
the on-resonance and off-resonance data are shown separately
(with the off-resonance
spectrum scaled to the on-resonance integrated luminosity
and corrected for the change in \ccbar cross-section).
Table~\ref{tab:pstar:spectrum-offpeak} shows the corresponding values.
The \ccbar peak is present
in both samples, but the \BB peak is only present in the
on-resonance sample. 
  Assuming baryon number conservation, the kinematic limit
  for $\Xi_c^0$ produced in the decays of $B$ mesons at
  \babar\ is $p^* = 2.135$~GeV$/c$.
We compare the on-resonance and scaled off-resonance samples
for $p^* \leq 2.15$~GeV$/c$ to obtain the yield of $\Xi_c^0$
produced in $B$ decays.
This is scaled by the number of $B$
mesons in the data sample (introducing a further 1.1\%
systematic uncertainty) to obtain:
\begin{eqnarray*}
  \mathcal{B}(B &\rightarrow& \Xi_c^0 X)
  \times \mathcal{B}(\Xi_c^0 \rightarrow \Xi^- \pi^+)
  \\ &=& (\iowaBresult \pm \iowaBstaterr \pm \iowaBsyserr) \times 10^{-4}
  .
\end{eqnarray*}
The yield of $\Xi_c^0$ produced in \ccbar events 
at an energy of 10.58~GeV is calculated from the
scaled off-resonance data set (for $p^* \leq 2.15$~GeV$/c$) and the
on-resonance data set (for $p^* > 2.15$~GeV$/c$). The yield is then
divided by the integrated luminosity (introducing a further 1.5\%
systematic uncertainty) to obtain the cross-section from the
continuum:
\begin{eqnarray*}
  \sigma(e^+ e^- &\rightarrow& \Xi_c^0 X)
  \times \mathcal{B}(\Xi_c^0 \rightarrow \Xi^- \pi^+)
  \\ &=& (\iowaCresult \pm \iowaCstaterr \pm \iowaCsyserr ) ~ \mathrm{fb}
  ,
\end{eqnarray*}
where both $\Xi_c^0$ and
$\overline{\Xi}_c^0$ are included in the cross-section.
The effect of initial state radiation is not isolated.

In summary, we have studied the $\Xi_c^0$ charmed baryon
at \babar\ through its decays to the $\Omega^- K^+$ and
$\Xi^- \pi^+$ final states using 116.1~\invfb of data.
The ratio of branching fractions of these decay modes was measured
to be $0.294 \pm 0.018 \pm 0.016$. This represents a substantial
improvement on the previous measurement~\cite{Henderson:1992cx} and is consistent
with a quark model prediction~\cite{Korner:1992wi}.
We have also measured the $p^*$ spectrum for $\Xi_c^0$ produced
at the $\Upsilon(4S)$ resonance.
The high rate
of $\Xi_c^0$ production at low $p^*$ in $B$ decays (below 1.2~GeV$/c$)
is particularly intriguing, implying that the invariant mass of the
recoiling antibaryon system is typically above 2.0~GeV$/c^2$.
This can be explained naturally by a substantial rate of charmed baryon pair
production through the $b \rightarrow c \bar{c} s$ weak decay
process~\cite{Ball:1990fw,Chernyak:1990ag,Sheikholeslami:1991fa,Dunietz:1994hj}
which was observed indirectly in a previous \babar\ analysis~\cite{Aubert:2004nb}.
In this Letter we measured the branching fraction product
$\iowaBranchingFractionProduct$ to be
$(\iowaBresult \pm \iowaBstaterr \pm \iowaBsyserr) \times 10^{-4}$;
the precision is significantly improved over the
previous measurement~\cite{Barish:1997pq}.
We have also measured the cross-section product 
$\iowaCrossSectionProduct$ from the continuum to be
$(\iowaCresult \pm \iowaCstaterr \pm \iowaCsyserr)$~fb.

\begin{acknowledgments}

%%\input acknow_PRL
%%%%%%%%%%%%%%%%%%%%%%% begin acknow_PRL.tex %%%%%%%%%%%%%%%%%%%%%%%
We are grateful for the excellent luminosity and machine conditions
provided by our \pep2\ colleagues, 
and for the substantial dedicated effort from
the computing organizations that support \babar.
The collaborating institutions wish to thank 
SLAC for its support and kind hospitality. 
This work is supported by
DOE
and NSF (USA),
NSERC (Canada),
IHEP (China),
CEA and
CNRS-IN2P3
(France),
BMBF and DFG
(Germany),
INFN (Italy),
FOM (The Netherlands),
NFR (Norway),
MIST (Russia), and
PPARC (United Kingdom). 
Individuals have received support from CONACyT (Mexico), A.~P.~Sloan Foundation, 
Research Corporation,
and Alexander von Humboldt Foundation.
%%%%%%%%%%%%%%%%%%%%%%%% end acknow_PRL.tex %%%%%%%%%%%%%%%%%%%%%%%%

\end{acknowledgments}

%% \bibliography{note}

\begin{thebibliography}{19}

\bibitem{ref:PDGbook}
Particle Data Group, 
S. Eidelman {\em et al.}, Phys.\ Lett.\ B {\bf 592}, 1 (2004).

\bibitem{footnote:cc}
Charge conjugate reactions are implied throughout.

\bibitem{Henderson:1992cx}
CLEO Collaboration,
S.~Henderson {\em et al.}, Phys.\ Lett.\ B {\bf 283}, 161 (1992). 

\bibitem{footnote:errors}
Throughout this Letter, the first uncertainty is statistical
and the second is systematic.

\bibitem{Korner:1992wi}
J.~G.~K\"{o}rner and M.~Kr\"{a}mer,
Z.\ Phys.\ C {\bf 55}, 659 (1992).


\bibitem{Ball:1990fw}
P.~Ball and H.~G.~Dosch, 
Z.\ Phys.\ C {\bf 51}, 445 (1991).

\bibitem{Chernyak:1990ag}
V.~L.~Chernyak and I.~R.~Zhitnitsky,
Nucl.\ Phys.\ {\bf B345}, 137 (1990).

\bibitem{Sheikholeslami:1991fa}
S.~M.~Sheikholeslami and M.~P.~Khanna,
Phys.\ Rev.\ D {\bf 44}, 770 (1991).

\bibitem{Dunietz:1994hj}
I.~Dunietz {\em et al.},
Phys.\ Rev.\ Lett.\ {\bf 73}, 1075 (1994).


\bibitem{Barish:1997pq}
CLEO Collaboration, 
B.~Barish {\em et al.}, Phys.\ Rev.\ Lett.\ {\bf 79}, 3599 (1997).

\bibitem{ref:babar}
\babar\ Collaboration,
B.\ Aubert {\em et al.},
%Nucl.\ Instrum.\ Methods A {\bf 479}, 1 (2002).
Nucl.\ Instr.\ Methods Phys.\ Res., Sect.\ A {\bf 479}, 1 (2002).

\bibitem{footnote:flightcuts}
For comparison, the mean transverse flight distances at \babar\ for
$\Xi^-$ and $\Omega^-$ which are produced in these $\Xi_c^0$
decays in \ccbar events are 5.3~cm and 2.3~cm, respectively, and
the mean flight distance of $\Lambda$ in these decays is 11.1~cm.
The corresponding resolutions are 0.7~mm, 0.7~mm, and 0.8~mm, respectively.

\bibitem{ref:pythia}
T.~Sjostrand {\em et al.}
Comput.\ Phys.\ Commun.\  {\bf 135}, 238 (2001).

\bibitem{ref:geant4}
S.~Agostinelli {\em et al.},
%Nucl.\ Instrum.\ Methods A {\bf 506}, 250 (2003).
Nucl.\ Instr.\ Methods Phys.\ Res., Sect.\ A {\bf 506}, 250 (2003).

\bibitem{Aubert:2004nb}
\babar\ Collaboration,
B.~Aubert {\em et al.},
Phys.\ Rev.\ D {\bf 70}, 091106 (2004).

\end{thebibliography}

\end{document}